\documentclass{article}

    \usepackage[preprint,nonatbib]{neurips_2020}
    
\usepackage[utf8]{inputenc} 
\usepackage[T1]{fontenc}    
\usepackage{hyperref}       
\usepackage{url}            
\usepackage{booktabs}       
\usepackage{amsfonts}       
\usepackage{nicefrac}       
\usepackage{microtype}      

\usepackage{amsmath}
\usepackage[square,numbers]{natbib}
\usepackage{graphicx}
\title{Angle Basis: a Generative Model and Decomposition for Functional Connectivity}

\author{%
  Anton Orlichenko\thanks{Corresponding author.}
  \\
  Department of Biomedical Engineering\\
  Tulane University\\
  New Orleans, LA 70118 \\
  \texttt{aorlichenko@tulane.edu} \\
  \And
  Gang Qu \\
  Department of Biomedical Engineering \\
  Tulane University\\
  New Orleans, LA 70118 \\
  \texttt{gqu1@tulane.edu} \\
  \And
  Ziyu Zhou \\
  Department of Computer Science \\
  Tulane University\\
  New Orleans, LA 70118 \\
  \texttt{zzhou11@tulane.edu} \\
  \And
  Zhengming Ding \\
  Department of Computer Science \\
  Tulane University\\
  New Orleans, LA 70118 \\
  \texttt{zding1@tulane.edu} \\
  \And
  Yu-Ping Wang \\
  Department of Biomedical Engineering \\
  Tulane University\\
  New Orleans, LA 70118 \\
  \texttt{wyp@tulane.edu} \\
}

\begin{document}

\maketitle

\begin{abstract}
  Functional connectivity (FC) is one of the most common inputs to fMRI-based predictive models, due to a combination of its simplicity and robustness. However, there may be a lack of theoretical models for the generation of FC. In this work, we present a straightforward decomposition of FC into a set of basis states of sine waves with an additional jitter component. We show that the decomposition matches the predictive ability of FC after including 5-10 bases. We also find that both the decomposition and its residual have approximately equal predictive value, and when combined into an ensemble, exceed the AUC of FC-based prediction by up to 5\%. Additionally, we find the residual can be used for subject fingerprinting, with 97.3\% same-subject, different-scan identifiability, compared to 62.5\% for FC. Unlike PCA or Factor Analysis methods, our method does not require knowledge of a population to perform its decomposition; a single subject is enough. Our decomposition of FC into two equally-predictive components may lead to a novel appreciation of group differences in patient populations. Additionally, we generate synthetic patient FC based on user-specified characteristics such as age, sex, and disease diagnosis. By creating synthetic datasets or augmentations we may reduce the high financial burden associated with fMRI data acquisition.
\end{abstract}

\section{Introduction and Related Work}

Functional magnetic resonance (fMRI) uses the blood oxygen level-dependent (BOLD) signal to map blood flow in the brain \cite{Belliveau1991FunctionalMO}, which is correlated to neural activity through the hemodynamic response function \cite{Rangaprakash2021-uf}. Early work in fMRI utilized statistical techniques on voxel intesities to measure brain region activation and group differences \cite{Monti2011-js}. More recently, functional connectivity (FC), the temporal Pearson correlation of BOLD signal across different regions of the brain, has been used to characterize brain state \cite{doi:10.1073/pnas.0135058100}. Regions are delineated using either a template \cite{Power2011FunctionalNO} or ICA \cite{Calhoun2009-kr}.

FC has recently been used for the prediction of subject age \cite{Li2018}, sex \cite{Zhang2020GenderDA}, race \cite{Orlichenko2023-wl}, or schizophrenia (SZ) diagnosis \cite{Li2020-ej}, at levels that in some cases meet or exceed prediction using genetic information \cite{Bracher-Smith2021-yb}. It has also been used to flag possible pre-clinical Alzheimer's disease by predicting a subject's brain age based on FC and comparing with actual age \cite{MILLAR2022119228}. Interestingly, it has even been used for mechanistic studies of aggression in response to olfactory stimulus \cite{Mishor2021-ma}, the traditional province of statistical fMRI techniques. In the hospital setting, it may be used in evaluating the severity and monitoring the treatment of concussions \cite{Kaushal2019-rw}. Dynamic FC, using either window-based or non-window-based \cite{Bolton2017-my} methods, has also been explored in the literature, with some studies showing benefit over static FC \cite{Rashid2016-ct} and others disagreeing \cite{Laumann2017-qo}.

Many studies have explored alternative measures of connectivity, such precision matrix-based (partial) correlation \cite{Kim2015-ev}, distance correlation \cite{Geerligs2016-ky}\cite{Luo2021-dq}, effective connectivity \cite{Stephan2010-os}, or phase lock value (PLV) \cite{Glerean2012-sw}. In general, these methods have been found to be largely similar \cite{Ahmadi2023-rk} in terms of usefulness. For example, partial correlation may have better test-retest reliability, but lose the meaning of region-region synchronization inherent in FC \cite{Mahadevan2021-fx}. \citet{Geerligs2016-ky} found that Pearson correlation has slightly worse results compared to distance correlation. In fact, FC and distance correlation-based connectivity maps are largely indistiguishable \cite{Luo2021-dq}. BrainNetCNN \cite{Kawahara2017-wp}, by contrast, works directly on time series data, but the resulting convolutional filters are not as interpretable as FC maps. Since FC is an established method, many recent papers focus on novel approaches. It should be noted, however, that even since the days of ANOVA-based statistical methods, fMRI has suffered from a reproducibility crisis \cite{Bennett2010-gy}\cite{Eklund2016-zl}. Therefore a focus on simplicity and reproducibility \cite{Turner2018-vq} may be preferred.

In terms of generative models which aim to simulate brain dynamics, the main focus has been on structural equation modeling (SEM) or dynamic causal modeling (DCM), used to generate effective connectivity \cite{Lin2009-fx}. Unfortunately, DCM has a high computational complexity that makes it unsuitable for working with a large number of regions \cite{Hidalgo-Lopez2021-sn}\cite{Sadeghi2020-wr}. Additionally, even the highest resolution fMRI only identifies the averaged activity of many individual neurons, further blurred by the hemodynamic response function \cite{Rangaprakash2021-uf}. Graphical models use FC, or some other connectivity method, as the basis for determining graph edge weights and then use thresholding to binarize these edges \cite{Kruschwitz2015-ff}. In effect, FC is the foundation on which most such methods are built. 

\citet{Zhao2020-xa} and others \cite{Liu2020-xg} have used GANs to help discriminate between patients and healthy controls. \citet{Tan2022-wp} have used a manifold-regularized Wasserstein distance GAN to achieve modest prediction gains. However, these GAN-based methods fail to model the underlying time-varying BOLD signal that underpins FC. Therefore they cannot be used for decomposition, dimensionality reduction, or easy interpretation as can our work.

In this work, we provide a theoretical basis for decomposition of FC based on regions of in-phase or out-of-phase sine waves, representing the bandpass-filtered BOLD signal at a region of interest (ROI). In conceptual terms, it has some similarity to the PLV determined using the Hilbert transform \cite{Glerean2012-sw}. The PLV, however, is not widely used in predictive models \cite{9721204}, and our own validation experiments have shown it performs much worse than other measures of connectivity when used for prediction tasks (see Supplemental Materials). By combining a sine-wave decomposition with a jitter value at each ROI, we are able to reconstruct the majority of FC, creating both a useful decompostion and a useful residual. Typical MRI studies require \$500-\$1000 per subject for data acquisition \cite{Szucs2020-hs}. Using our model for the generation of synthetic FC based on specific phenotype settings, we may alleviate some of the high financial burden of fMRI data acquisition, allowing greater accessibility to this modality. 

\section{Methods}


We describe angle basis as an approximation of the Fourier series representation of the time domain BOLD fMRI signal $x(t)$, from which FC is derived. We motivate the underlying model by a connection to the well-known but less widely used PLV. Note that the FC is defined as the Pearson correlation of the time domain signals between regions $c$ and $d$:

\begin{equation}
    \rho_{cd} = \frac{\sigma_{cd}^2}{\sqrt{\sigma_c^2\sigma_d^2}}
\end{equation}

\subsection{Phase Lock Value (PLV)}

Any time domain signal can be represented in a basis of complex sinusoids via the Fourier transform:

\begin{equation}
    X(\omega) = \int_{-\infty}^\infty x(t)e^{-i{\omega}t}dt
\end{equation}

The Hilbert transform, acting on the time domain signal, imparts a $\pi/2$ phase shift to every complex sinusoid depending on the sign of the sinusoid frequency:

\begin{equation}
    \text{H}[x](t) = \frac{1}{\pi}\text{p.v.}\int_{-\infty}^{\infty}\frac{x(\tau)}{t-\tau}d\tau
\end{equation}

Here p.v. represents the Cauchy principal value of the improper integral. Using the Hilbert transform, one may change the original signal into an analytic signal $x_a(t)$ containing only positive frequencies.

\begin{equation}
    x_a(t) = x(t)+i\text{H}[x(t)]
\end{equation}

The analytic signal may be factorized into a low-frequency amplitude envelope $a(t)$ and a phase term $\theta(t)$, so long as Bedrosian's theorem \cite{Brown1986-qz} holds, i.e. as long as the support of $a(t)$ and $\theta(t)$ are different. This may be the case for fMRI signals, which are typically bandpass filtered.

\begin{equation}
    x_a(t) = a(t)e^{i\theta(t)}
\end{equation}

The connectivity between two regions $c,d$ may be quantified as the PLV, a measure of the dispersion of phase differences between the two regions \cite{Glerean2012-sw}.

\begin{equation}
    \begin{split}
        \theta_{cd}(t) &= \theta_c(t)-\theta_d(t) \\
        \text{PLV}_{cd} &= \frac{1}{T}\left\lvert\sum_{t=1}^{T}e^{i\theta_{cd}(t)}\right\rvert
    \end{split}
\end{equation}

Although PLV has been widely used in neuroscience \cite{Lachaux1999-lg}, it is dependent on the passband and has relatively poor predictive ability (see Supplemental Materials). Additionally, one may wish to work directly with the FC. Given this motivation, we present a simplified but more robust model.

\begin{figure}
    \centering
    \hspace*{-1cm}\includegraphics[width=16cm]{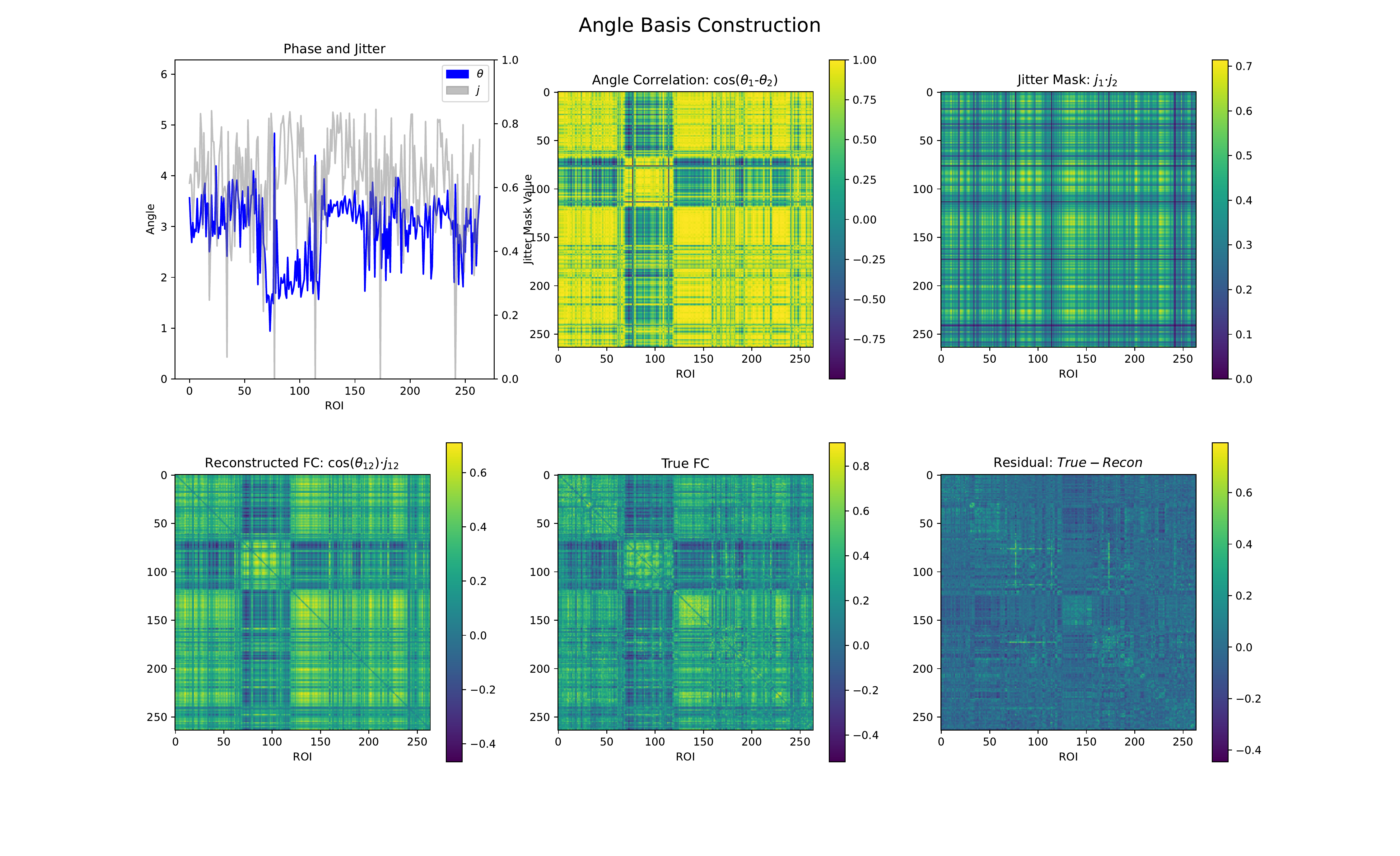}
    \caption{Example of the angle basis reconstruction of one representative subject from the PNC dataset. The prominent block corresponds to the default mode network (DMN). The scan is probably of a female, due to the high within-DMN connectivity \cite{Ficek-Tani2022-yw}. Here only one basis is used.}
    \label{fig:abonesubject}
\end{figure}

\subsection{Angle Basis Decomposition}

First, we represent the signal as a Fourier series, assuming, as in the case of the PLV, that the amplitude envelope of the passband-filtered signal is constant with respect to the phase signal.

\begin{equation}
    x(t) = \sum_{n=1}^{N}A_n\text{sin}({\omega_n}t+\theta_n)
\end{equation}

 In the simplest case, we approximate the BOLD signal at each region $c$ as a single sine wave with arbitrary constant amplitude and phase. 

\begin{equation}
    \hat{x}_c(t) = A_c\text{sin}({\omega}t+\theta_c)
\end{equation}

The correlation between ROIs $c$ and $d$ is then simply the cosine of the phase difference. 

\begin{equation}
    \hat{\rho}_{cd} = \text{cos}(\theta_c-\theta_d)
\end{equation}

We reduce the number of frequencies required to decompose the signal by introducing a jitter component, which models the signal at a region "jittering" such that it becomes uncorrelated with all other regions. 

\begin{equation}
    \hat{\rho}_{cd}^{(n)} = j_{c}^{(n)}{\cdot}j_{d}^{(n)} \cdot\text{cos}(\theta_c^{(n)}-\theta_d^{(n)})
\end{equation}

Here the superscript $(n)$ refers to the correlation in the $n$\textsuperscript{th} basis. Each scan can be modeled with up to $N$ bases, where the bases represent either different temporal regions or orthogonal sinusoids of different frequencies. We then calculate the overall correlation between regions $c,d$ as follows:

\begin{equation}
    \tilde{\rho}_{cd} = \frac{1}{N}\sum_{n=1}^{N}\hat{\rho}_{cd}^{(n)}
\end{equation}

We understand this formula is only valid when amplitudes are the same across all basis sinusoids; however we find this model to be useful regardless. Additionally, differences in window lengths or amplitudes can be incorporated into the $j_c^{(n)}\cdot{j_d^{(n)}}$ term in a uniform way. 


The decomposition is estimated via gradient descent by minimizing the RMSE between the reconstructed and true FC, i.e. the residual $r_{cd}$.

\begin{equation}
\begin{split}
    &r_{cd} = \rho_{cd}-\tilde{\rho}_{cd} \\
    \underset{\mathbf{\theta}, \mathbf{j}}{\text{min}} \quad \frac{1}{R^2}\sum_{cd}r_{cd}^2
    &\qquad \text{s.t.} \quad 0\leq\theta_c^{(n)}<2\pi, 
    \quad 0{\leq}j_c^{(n)}\leq1 
\end{split}
\end{equation}

Here $R$ is the number of regions in the remplate, $\rho_{cd},\tilde{\rho}_{cd}$ are the true and estimated FC values, respectively, and $r_{cd}$ is the residual. Values of jitter that become less than 0 or greater than 1 are clipped to those values while phase angles wrap around.

Figure~\ref{fig:abonesubject} shows the decomposition of a representative's subject's FC into a single basis plus residual. The template we used used contains 264 regions, meaning that each basis must contain 528 values for phase and jitter. In section \ref{subsec:nbaseseffect}, we show that 5-10 bases are enough to recreate the predictive capacity of FC. In contrast, there are 34,716 non-duplicate values in a subject's FC.

Angle basis estimation is carried out using PyTorch \cite{DBLP:journals/corr/abs-1912-01703} with CUDA support.\footnote{\url{https://pytorch.org/}} Our source code may be found at \url{https://github.com/aorliche/AngleBasis}. Training time is approximately one second per subject for a one basis decomposition, up to 10 seconds for higher numbers of basis sinusoids, using a single GPU.

\section{Experiments}
\label{subsec:experiments}

We validate the usefulness of the decomposition in five downstream tasks:

\begin{itemize}
    \item Identification of the same subject from different scans
    \item Prediction of age, sex, race, and schizophrenia diagnosis across two different datasets
    \item Transfer of models trained on one dataset to predict sex or race to another dataset
    \item Prediction using reduced feature numbers (feature selection)
    \item Generation of synthetic FC
\end{itemize}

Prediction was carried out using vectorizations of the true FC matrix, angle basis reconstruction, angle basis residual, or ensemble of angle basis and residual. In addition, a Factor Analysis was used to create latents and residuals for the identifiability task and age prediction. Identifiability was carried out by cosine similarity, which was found to give better results than Euclidean distance for all inputs tested. A deep autoencoder (AE) with 2 hidden layers of 2000 and 2000 neurons with ReLU activation was also used to create a reconstruction/residual for the prediction tasks.

The models used for prediction were simple Ridge and Logistic Regression models, from the scikit-learn implementation\cite{scikit-learn},\footnote{\url{https://scikit-learn.org/stable/}}  with a maximum of 1000 iterations and regularization parameters $\alpha=1$ and $C=1$, respectively. Models were evaluated using an 80/20 train test split over 20 bootstrap repetitions. When using Ridge Regression (age prediction), both training and test sets were centered using the mean age of the training set. For race prediction, only African Ancestry (AA) and European Ancestry (EA) ethnicities were used, because our datasets were enriched for these groups, with other ethnicities making up less than 10\% of subjects. Although FC-based predictions are not always improved by using more expressive models \cite{Pervaiz2020-mf}, we include a comparison with an MLP and BrainNetCNN \cite{Kawahara2017-wp} in the Supplemental Materials.

\subsection{Datasets}

We tested our model on two different, large, widely-used datasets available to researchers via application. Both datasets were pre-processed with SPM12,\footnote{\url{https://www.fil.ion.ucl.ac.uk/spm/software/spm12/}} including intra-subject co-registration, warping to MNI space, smoothing with a 5mm FWHM kernel, temporal bandpass filtering from 0.01 to 0.15 Hz, and time series extraction using the 264-region Power atlas \cite{Power2011FunctionalNO}. Power atlas regions are discrete balls of 5mm radius from which average BOLD signal was extracted.

\subsubsection{Philadelphia Neurodevelopmental Cohort (PNC)}

The PNC dataset is a large, widely used dataset consisting of children and young adults 8-22 years old, 1,529 of whom have fMRI scans \cite{Satterthwaite2014NeuroimagingOT} and over 9,000 of whom have genetic information \cite{Glessner2010-xg}. Most subjects have three scans, one for a resting state, one for a working memory \cite{Ragland2002-si}, and one for an emotion identification in-scanner task. A wide variety of phenotype information is available, including sex, race, cognitive battery, and questionnaire info \cite{Calkins2015-su}. Subjects were not specifically selected due to prior neurological problems. Of subjects having scans, 725 are male, 804 are female, 660 are AA ethnicity, 690 are EA ethnicity, and 179 are of other ethnicity. More detailed phenotype data and correlations are provided in the Supplemental Materials.

\subsubsection{Bipolar and Schizophrenia Network for Intermediate Phenotypes (BSNIP)}

The BSNIP dataset is a large multi-site fMRI dataset acquired to further understand schizophrenia, schizoaffective disorder, and bipolar disorder \cite{Tamminga2014-sq}. Our version of the dataset contains fMRI scans for 1,245 adult subjects acquired over 6 sites. These include 509 total patients (of which 199 have SZ diagnosis), 494 relatives of patients, and 242 normal controls. Subjects demographics are as follows: 528 are male, 717 are female, 387 are AA ethnicity, 778 are Caucasian (CA) ethnicity, and 81 are of other ethnicity. See Supplemental Materials for additional phenotype data. 


\subsection{Identification of Individuals}

\begin{figure}
    \centering
    \hspace*{-1cm}\includegraphics[width=16cm]{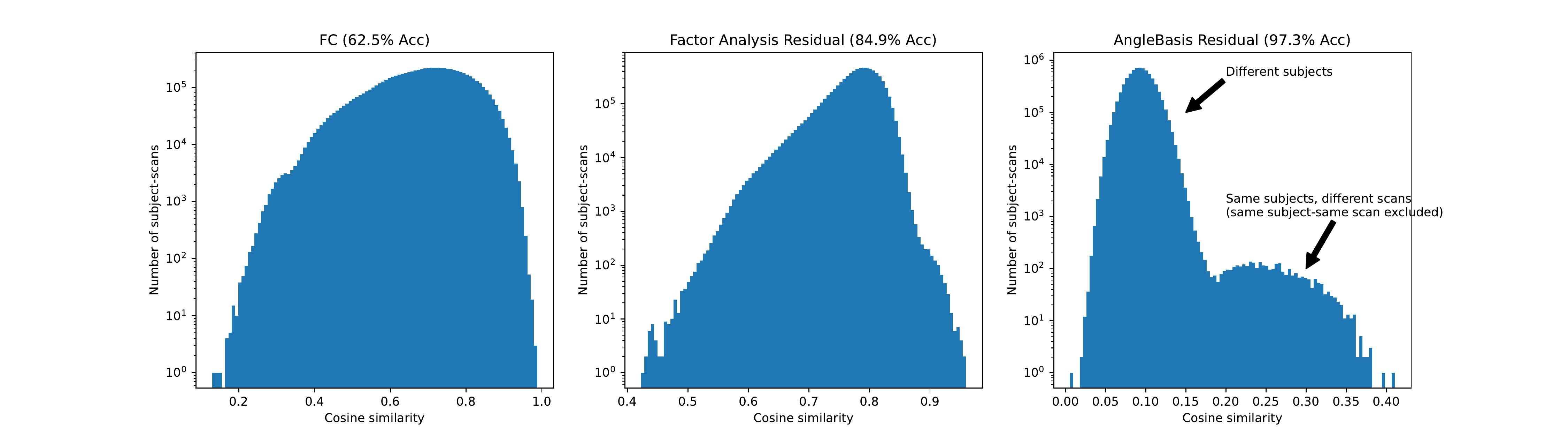}
    \caption{Identification of subjects from different scans in the PNC dataset. Histograms show the number of scan pairs at a particular cosine similarity value. Most subjects in the PNC dataset contain 3 scans with different in-scanner tasks. A total of 3849 scans were used.}
    \label{fig:fingerprint}
\end{figure}

We find that taking the residual of the angle basis reconstruction using one basis greatly increases the identification rate of subjects in the PNC dataset, from 62.5\% using FC to 97.3\% using the angle basis residual (see Figure~\ref{fig:fingerprint}). In addition, we find that a Factor Analysis residual (not PCA) with $N=10$ factors also increases identification rate to 84.9\%. Previous work has found similar different-task same-subject identification as our FC result \cite{Finn2015-ft}, as well as a slightly better result using a Factor Analysis-like method to produce a residual \cite{Cai2019-vp}. However, as seen in Figure~\ref{fig:ageres}, we find that the Factor Analysis method is not amenable to high prediction accuracy. Additionally, Factor Analysis, PCA, and the deep AE require knowledge of and have a dependence on the entire dataset. 

\subsection{Prediction Accuracy, Model Transfer, and Feature Selection}

\begin{figure}
    \centering
    \hspace*{-0.5cm}\includegraphics[width=13cm]{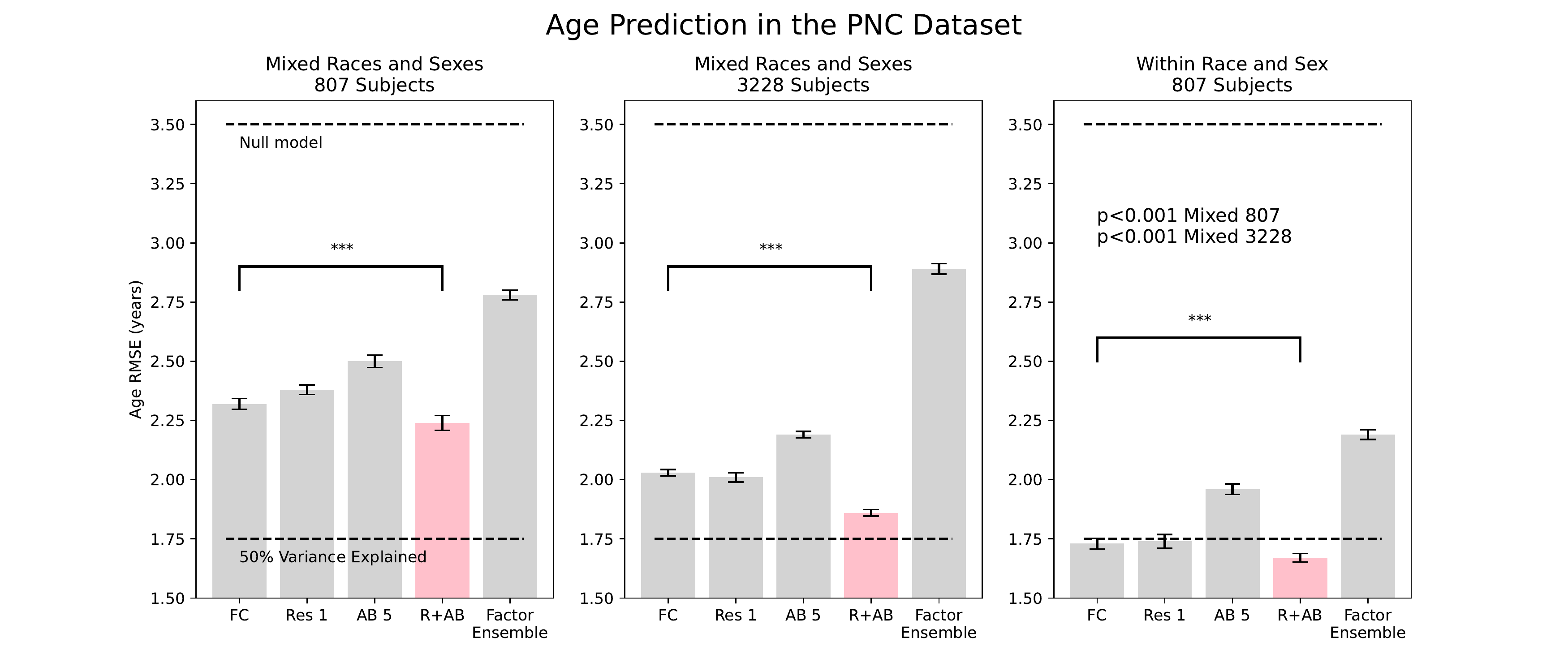}
    \caption{Age prediction in the PNC dataset. The ensemble of residual and angle basis is slightly but significantly better than the FC from which it was derived. Additionally, we find a strong effect for predicting within race and sex-matched compared to within mixed groups.}
    \label{fig:ageres}
\end{figure}

Figure~\ref{fig:ageres} shows that prediction RMSE of age in the PNC dataset is comparable between FC and the first residual, and slightly better for the ensemble of residual and angle basis reconstruction. In addition, we see that removing the confounding effects of sex and race results in much stronger age prediction accuracy. In fact, increasing the dataset by a factor of four still doesn't match the accuracy of predicting within sex and race-matched groups. We also note that the Factor Analysis method does not have strong predictive fidelity for age prediction, and the deep AE ensemble performs poorly in predictive tasks. PCA was similar to the deep AE (not shown). We additionally tested the prediction fidelity of sex and race in the PNC dataset; and sex, race, and schizophrenia diagnosis in the BSNIP dataset. The results are shown in Table~\ref{tab:results}. That the residual is highly useful for prediction is supported by related work manipulating PCA-based identifiability to enhance predictive performance \cite{Svaldi2021-pz}. 


\begin{table}
    \centering
    \caption{Classification Accuracy (AUC)}
    \begin{tabular}{cccccc}
        \toprule
        Dataset & Predictive Task & FC & Deep AE Ens & AB+Res Ens & p-value \\
        \midrule
        BSNIP & SZ/NC & $0.785\pm0.048$ & $0.701\pm0.062$ & $\mathbf{0.804\pm0.040}$ & 0.240 \\
        BSNIP & Sex & $0.755\pm0.023$ & $0.645\pm0.085$ & $\mathbf{0.791\pm0.022}$ & $0.002$ \\
        BSNIP & Race & $0.845\pm0.022$ & $0.739\pm0.051$ & $\mathbf{0.866\pm0.023}$ & $0.108$ \\
        PNC & Sex & $0.886\pm0.006$ & $0.744\pm0.065$ & $\mathbf{0.923\pm0.010}$ & $<0.001$ \\
        PNC & Race & $0.946\pm0.007$ & $0.812\pm0.040$ & $\mathbf{0.973\pm0.003}$ & $<0.001$ \\
        \midrule
        BSNIP$\rightarrow$PNC & Sex & $0.667\pm0.017$ & $0.601\pm0.032$ & $\mathbf{0.700\pm0.013}$ & $<0.001$ \\
         BSNIP$\rightarrow$PNC & Race & $0.807\pm0.018$ & $0.710\pm0.010$ & $\mathbf{0.847\pm0.012}$ & $<0.001$ \\
         PNC$\rightarrow$BSNIP & Sex & $0.629\pm0.019$ & $0.572\pm0.019$ & $\mathbf{0.667\pm0.013}$ & $<0.001$ \\
        PNC$\rightarrow$BSNIP & Race & $0.800\pm0.010$ & $0.702\pm0.022$ & $\mathbf{0.832\pm0.009}$ & $<0.001$ \\
        \bottomrule
    \end{tabular}
    \label{tab:results}
\end{table}


We next tested the transfer of sex and race prediction models from the PNC to the BSNIP dataset and vice versa. Models were trained wholly on one dataset and evaluated on the other. The results are shown in Table~\ref{tab:results}. It should be remembered that the PNC dataset is composed of ostensibly normal pre-teens, teens, and young adults, while the BSNIP dataset is composed of adults with schizophrenia-like diseases, their relatives, and normal controls. The high model transfer AUC for race prediction makes the case for biologically significant invariant features. Sex, by contrast, is less stable to model transfer, though prediction is still somewhat robust. It is well-known, for instance, that, on average, women demonstrate stronger within-DMN connectivity than men \cite{Ficek-Tani2022-yw}. 



We used both Ridge and Lasso models to estimate the most relevant features for age prediction in the PNC dataset. Features were estimated on 50\% of the dataset, with the remaining 50\% used in an 80/20 split to train on and validate the selected features. See Supplemental Materials for results. 


\subsection{Effect of the Number of Bases}
\label{subsec:nbaseseffect}

\begin{figure}
    \centering
    \hspace*{-1cm}\includegraphics[width=15cm]{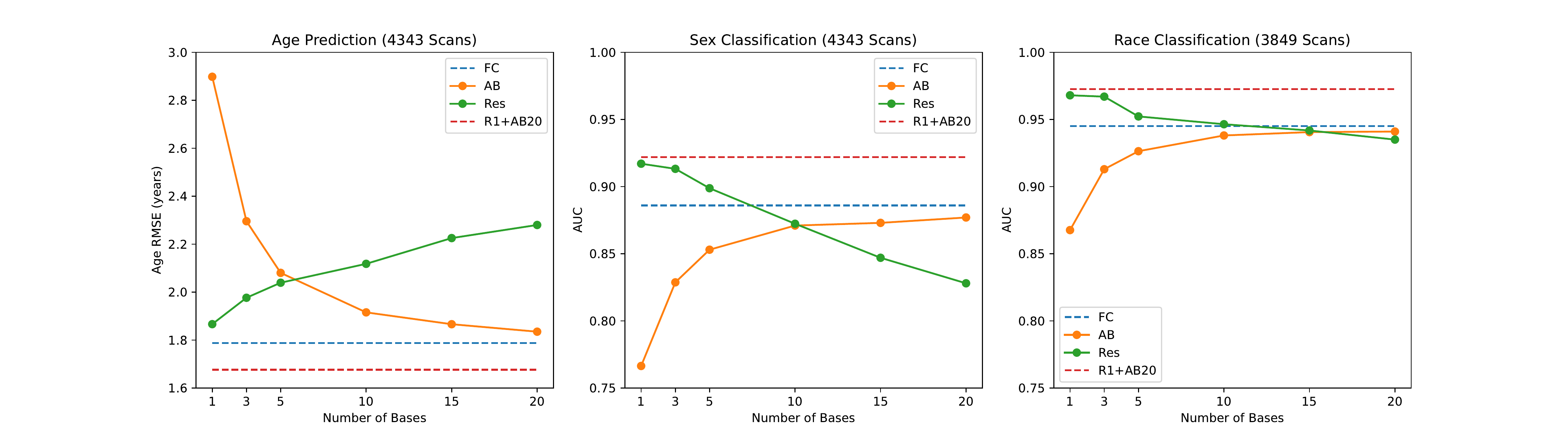}
    \hspace*{-1cm}\includegraphics[width=15cm]{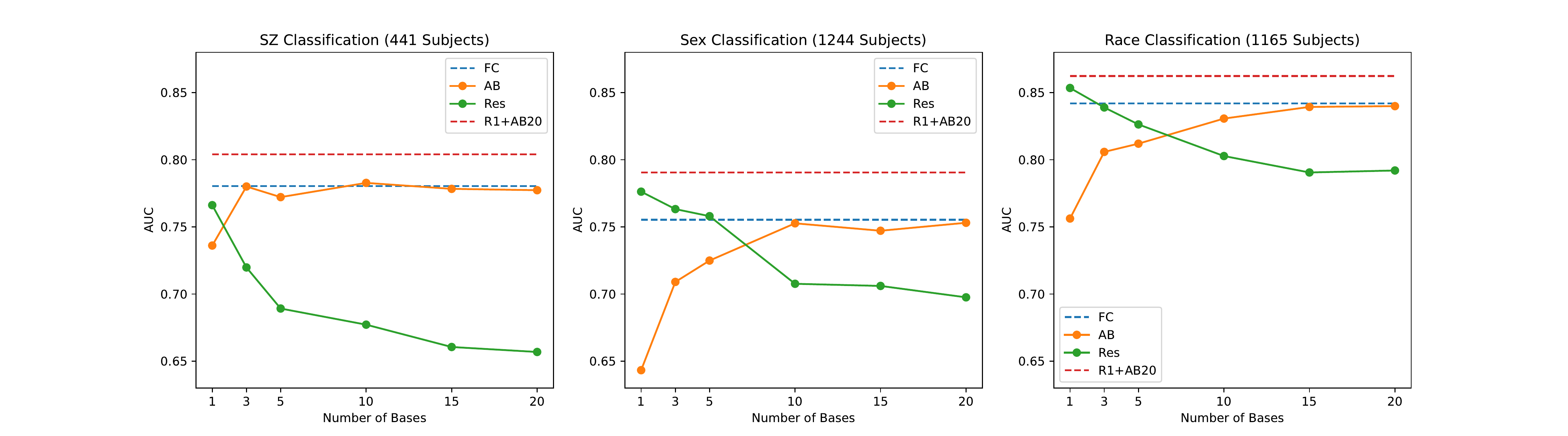}
    \caption{The prediction fidelity (RMSE and AUC) of the angle basis reconstruction and residual as a function of the number of bases. Top plot is for the PNC dataset, bottom is for the BSNIP dataset.}
    \label{fig:abnumber}
\end{figure}

We find that prediction accuracy increases monotonically with increasing number of bases (see Figure~\ref{fig:abnumber}), corresponding to decreasing reconstruction error. In contrast, prediction accuracy of the residual decreases monotonically with increasing number of bases. Using the first residual and the 20\textsuperscript{th} angle basis reconstruction in an ensemble results in a higher accuracy than FC. Using FC twice in an ensemble is not significantly different compared to a single FC-based prediction.

\subsection{Generation of Synthetic FC}

We trained a set of 20 predictive models for age, race, sex, and schizophrenia diagnosis, and used them to generate synthetic FC via the angle basis framework using random seeds. The results for a synthetic 30 year old normal CA female and synthetic 30 year old schizophrenic AA male are shown in Figure~\ref{fig:synthetic}. Additional examples and the procedure for generating synthetic FC are given in the Supplemental Materials.

\begin{figure}
    \centering
    \hspace*{-1.25cm}
    \vspace*{-0.75cm}
    \includegraphics[width=16cm]{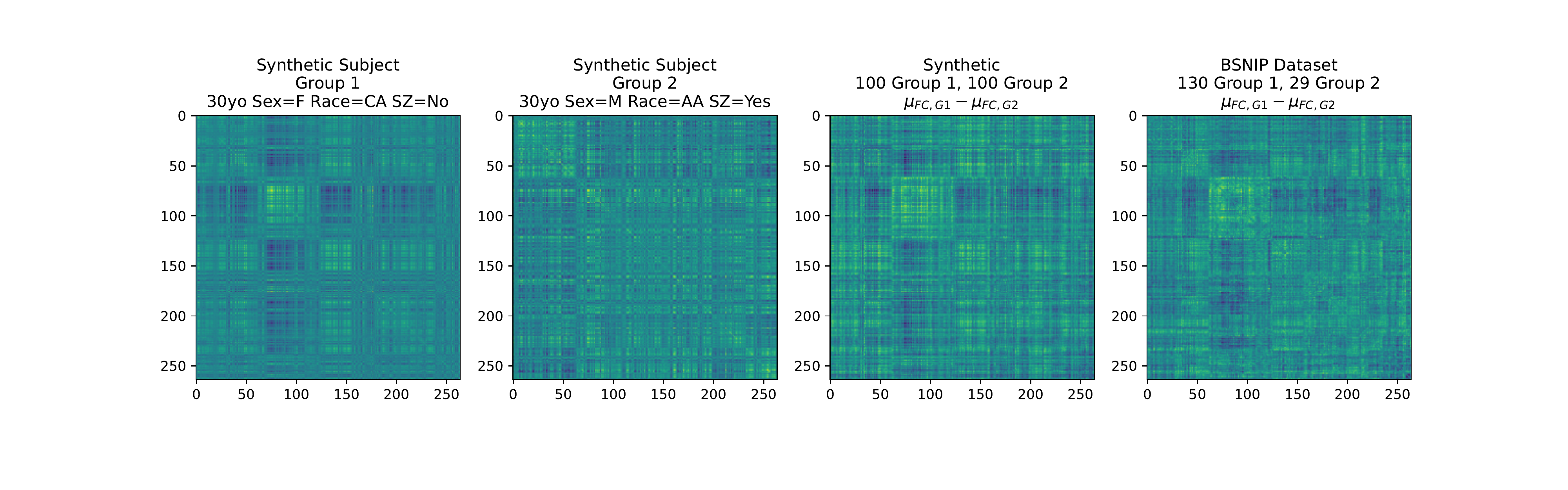}
    \caption{Generation of synthetic subjects that capture characteristics of patient FC. The synthetic female usually, but not always, has strong intra-DMN connectivity, while the synthetic man has a more flat connectivity profile. See the Supplemental Materials for additional examples and the distribution of synthetic scans for a given phenotype input.}
    \label{fig:synthetic}
\end{figure}

\section{Conclusion}

In this work, we present a generative model for and decomposition of FC, which requires only a single subject's connectivity matrix as input, and which generates a highly accurate reconstruction using a fraction of the original parameters. We find that both the reconstruction and residual have similar predictive ability, with an ensemble of the two surpassing FC in most cases. Additionally, we find that the residual is highly useful in subject identification or fingerprinting, achieving a nearly 100\% subject identification rate on different in-scanner task scans. The estimation of the decomposition is fast, taking seconds for a single subject using a workstation with a mid-priced GPU. Our goal is to use this work as a baseline for the creation of synthetic FC datasets that incorporate information about population characteristics, with the aim to reduce the financial burden of fMRI acquisition. 

\section{Acknowledgements}

The authors would like acknowledge the NIH (grants R01 GM109068, R01 MH104680, R01 MH107354, P20 GM103472, R01 REB020407, R01 EB006841, R56 MH124925) and NSF (\#1539067) for partial funding support.

fMRI and phenotype data came from the Neurodevelopmental Genomics: Trajectories of Complex Phenotypes database of genotypes and phenotypes repository, dbGaP Study Accession ID phs000607.v3.p2, as well as the Bipolar and Schizophrenia Network for Intermediate Phenotypes study (\url{http://b-snip.org/}).

\bibliography{anglebasis}

\appendix

\section{Source Code}

Code is available at \url{https://github.com/aorliche/AngleBasis}. Data used in our experiments is only available by request to the appropriate agencies, due to patient privacy issues (see Acknowledgements section in main paper). However, we provide sample preprocessed data taken from \url{https://openneuro.org/} in the GitHub repository. We also provide frozen models and Jupyter notebook code for the generation of synthetic FC based on user-provided subject characteristics.

\section{Effect of Jitter vs Angle on Decomposition}

We investigated the possibility of using only jitter or only angle as a basis for decomposing FC. Although jitter could be estimated using gradient descent with constant phase angle, a more efficient way is to use the truncated eigendecomposition of the FC. For example, a rank-5 decomposition would be created as follows:

\begin{equation}
    \begin{split}
        \mathbf{X} &= \mathbf{V}\mathbf{\Lambda} \mathbf{V}^{\text{T}} \\
        \hat{\lambda}_{ii} &= 
        \begin{cases}
            \lambda_{ii}, &\quad i\leq5 \\
            0, &\quad \text{otherwise}
        \end{cases} \\
        \hat{\mathbf{X}} &= \mathbf{V}\hat{\mathbf{\Lambda}}\mathbf{V}^{\text{T}}
    \end{split}
\end{equation}

\noindent where $\mathbf{X}$ is a subject's FC, $\hat{\mathbf{X}}$ is the low-rank estimate, and $\mathbf{\Lambda}$ is the ordered diagonal matrix of eigenvalues. This is possible because the FC is a positive semidefinite matrix.

\begin{figure}
    \centering
    \hspace*{-0.5cm}\includegraphics[width=14cm]{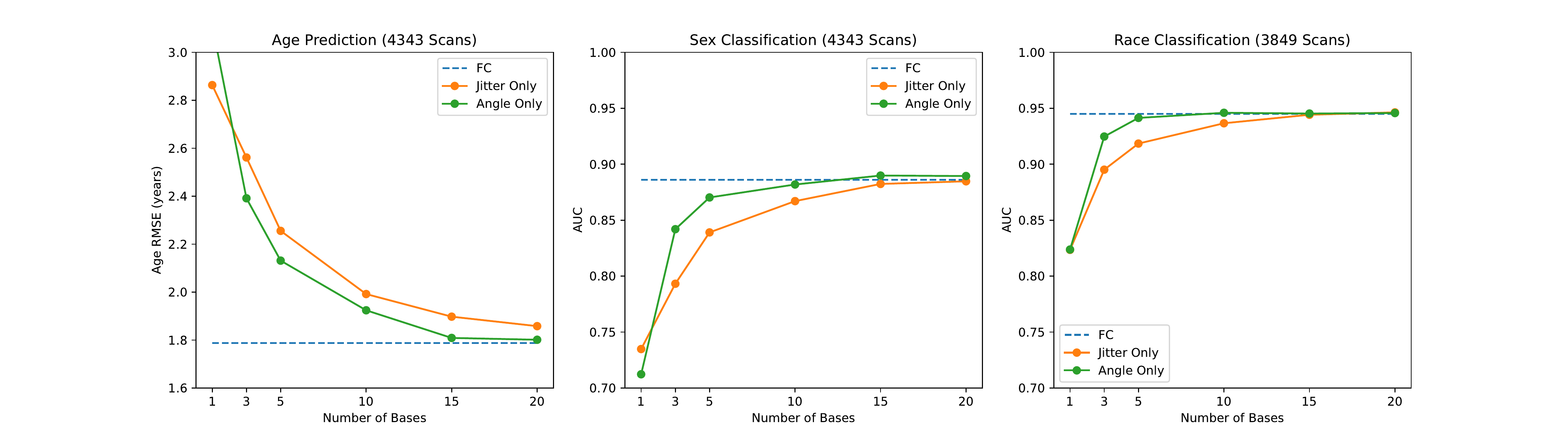}
    \caption{Prediction accuracy in the PNC dataset using only jitter vs using only the angle basis (cosine of phase difference).}
    \label{fig:abvseigs}
\end{figure}

Angle basis decomposition without jitter was estimated using gradient descent. The predictive accuracy of both models is shown in Figure~\ref{fig:abvseigs}.

\section{Comparison of Linear and Nonlinear Models for Downstream Tasks}

\begin{table}[]
    \centering
    \caption{Comparison of Linear and Deep Models in BSNIP Dataset}
    \vspace{0.3cm}
    \begin{tabular}{lcc}
        \toprule
        \multicolumn{3}{l}{Schizophrenia Diagnosis} \\
        \midrule
         Model & AUC & p-value \\
         Logistic Regression & $\mathbf{0.785\pm0.048}$ & - \\
         MLP & $0.771\pm0.025$ & 0.128 \\
         BrainNetCNN & $0.723\pm0.028$ & $<0.001$ \\
         \bottomrule
    \end{tabular}
    \begin{tabular}{lcc}
        & & \\
        \toprule
        \multicolumn{3}{l}{Sex Prediction} \\
        \midrule
         Model & AUC & p-value \\
         Logistic Regression & $\mathbf{0.755\pm0.022}$ & - \\
         MLP & $0.753\pm0.012$ & 0.90 \\
         BrainNetCNN & $0.731\pm0.049$ & 0.069 \\
         \bottomrule
    \end{tabular}
    \begin{tabular}{lcc}
        & & \\
        \toprule
        \multicolumn{3}{l}{Race Prediction} \\
        \midrule
         Model & AUC & p-value \\
         Logistic Regression & $\mathbf{0.849\pm0.011}$ & - \\
         MLP & $\mathbf{0.849\pm0.009}$ & - \\
         BrainNetCNN & $0.755\pm0.034$ & $<0.001$ \\
         \bottomrule
    \end{tabular}
    \label{tab:deepmodels}
\end{table}

We make a comparison between Logistic Regression, an MLP model, and the BrainNetCNN, similar to \citet{Pervaiz2020-mf}. As in that work, we find the linear model performed the best, with BrainNetCNN having worse performance (see Table~\ref{tab:deepmodels}). The small number of subjects and high inter-subject variability may hinder the advantage of more expressive models.

\section{Prediction using the Phase Lock Value Directly}

\begin{figure}
    \centering
    \hspace*{-1cm}
    \includegraphics[width=14cm]{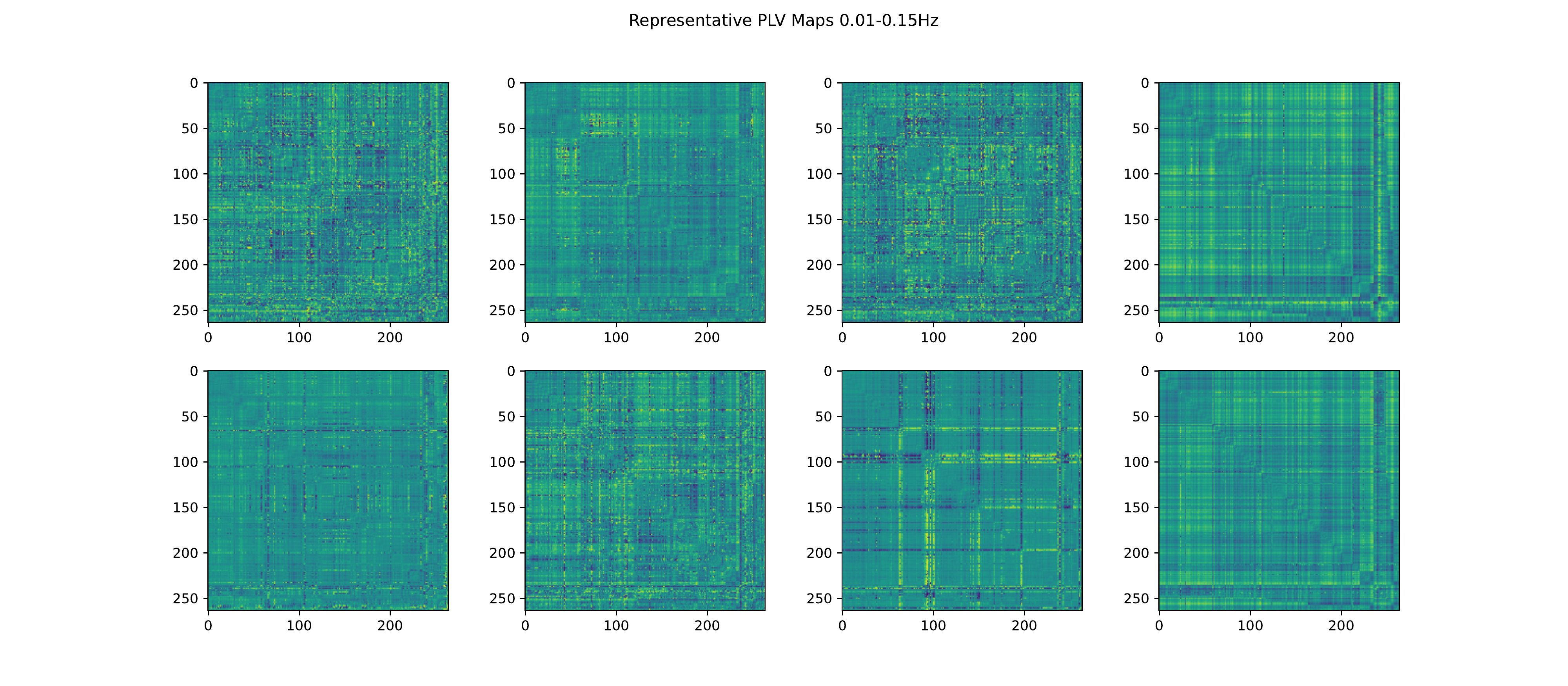}
    \caption{Phase lock value-based connectivities from the PNC dataset. Each image represents one subject.}
    \label{fig:plv}
\end{figure}

We include in Figure~\ref{fig:plv} a set of representative PLV connectivities from the PNC dataset. As seen in Table~\ref{tab:plv}, the PLV connectivies themselves are poor predictors of phenotypes.

\begin{table}
    \centering
    \caption{PLV Prediction Accuracy in the PNC Dataset}
    \vspace{0.3cm}
    \begin{tabular}{ccccc}
        \toprule
        Metric & Predictive Task & FC & PLV & Null Model \\
        \midrule
        RMSE (years) & Age & $\mathbf{1.79\pm0.02}$ & $3.13\pm0.07$ & $3.5$ \\
        AUC & Sex & $\mathbf{0.886\pm0.006}$ & $0.605\pm0.046$ & $0.53$ \\
        AUC & Race & $\mathbf{0.946\pm0.007}$ & $0.704\pm0.033$ & $0.52$ \\
        \bottomrule
    \end{tabular}
    \label{tab:plv}
\end{table}

\section{Additional Demographic Information}

\begin{figure}
    \centering
    \includegraphics[width=6cm]{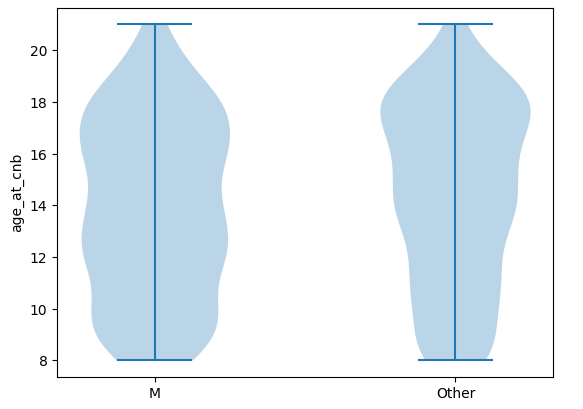}
    \includegraphics[width=6cm]{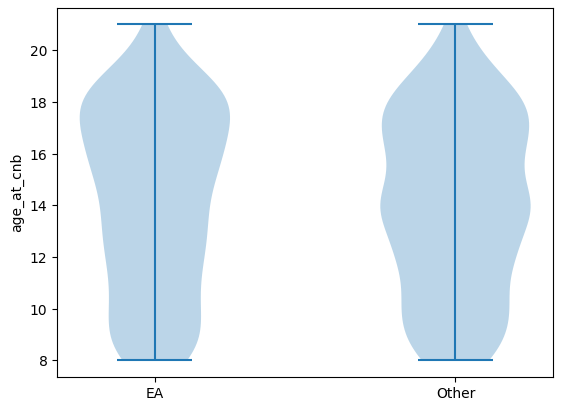}
    \caption{Sex and ethnicity broken down by age in the PNC dataset. EA is European Ancestry, Other is African Ancestry.}
    \label{fig:pnc}
\end{figure}

\begin{figure}
    \centering
    \includegraphics[width=6cm]{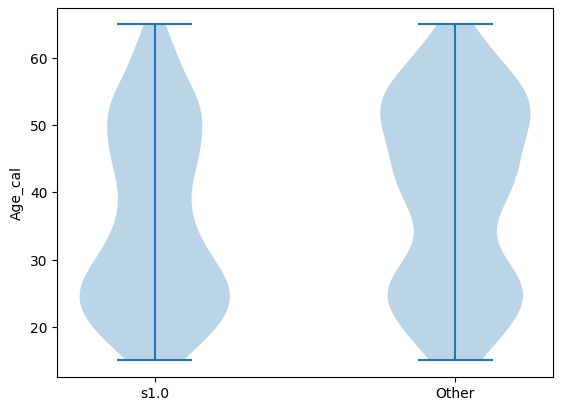}
    \includegraphics[width=6cm]{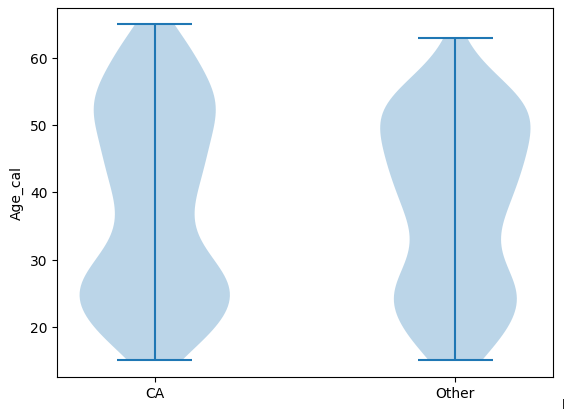}
    \caption{Sex and ethnicity broken down by age in the BSNIP dataset. ``s1.0" refers to male and ``Other" refers to female. CA is Caucasian Ancestry, Other is African Ancestry.}
    \label{fig:bsnip}
\end{figure}

\begin{figure}
    \centering
    \includegraphics[width=6cm]{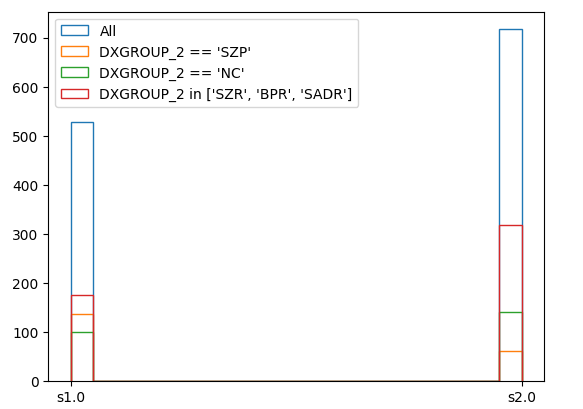}    \includegraphics[width=6cm]{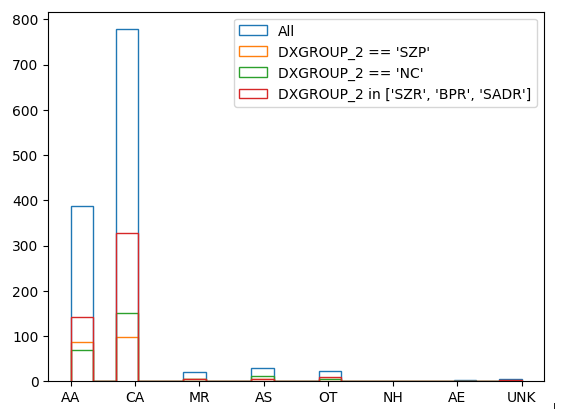}
    \caption{Sex and ethnicity broken down by patient subtype in the BSNIP dataset. ``s1.0" refers to male, ``s2.0" refers to female. Besides normal controls (NC) and patients (SZP, BPP, and SADP), there are also relatives of patients (SZR, BSPR, and SADR).}
    \label{fig:bsnip_more}
\end{figure}

We include more detailed demographic information on the PNC and BSNIP datasets in Figures~\ref{fig:pnc}, \ref{fig:bsnip} and \ref{fig:bsnip_more}.

\section{Generation of Synthetic FC}

Synthetic FC was generated by fitting 20 linear predictive models to the BSNIP dataset. The models used were Ridge Regression for age prediction and Logistic Regression for sex, race, and schizophrenia diagnosis prediction. Hyperparameters were the same as in the main paper. Models were fit on $20\%$ random samples of the dataset. 

\begin{figure}
    \centering
    \hspace*{-1cm}
    \includegraphics[width=14cm]{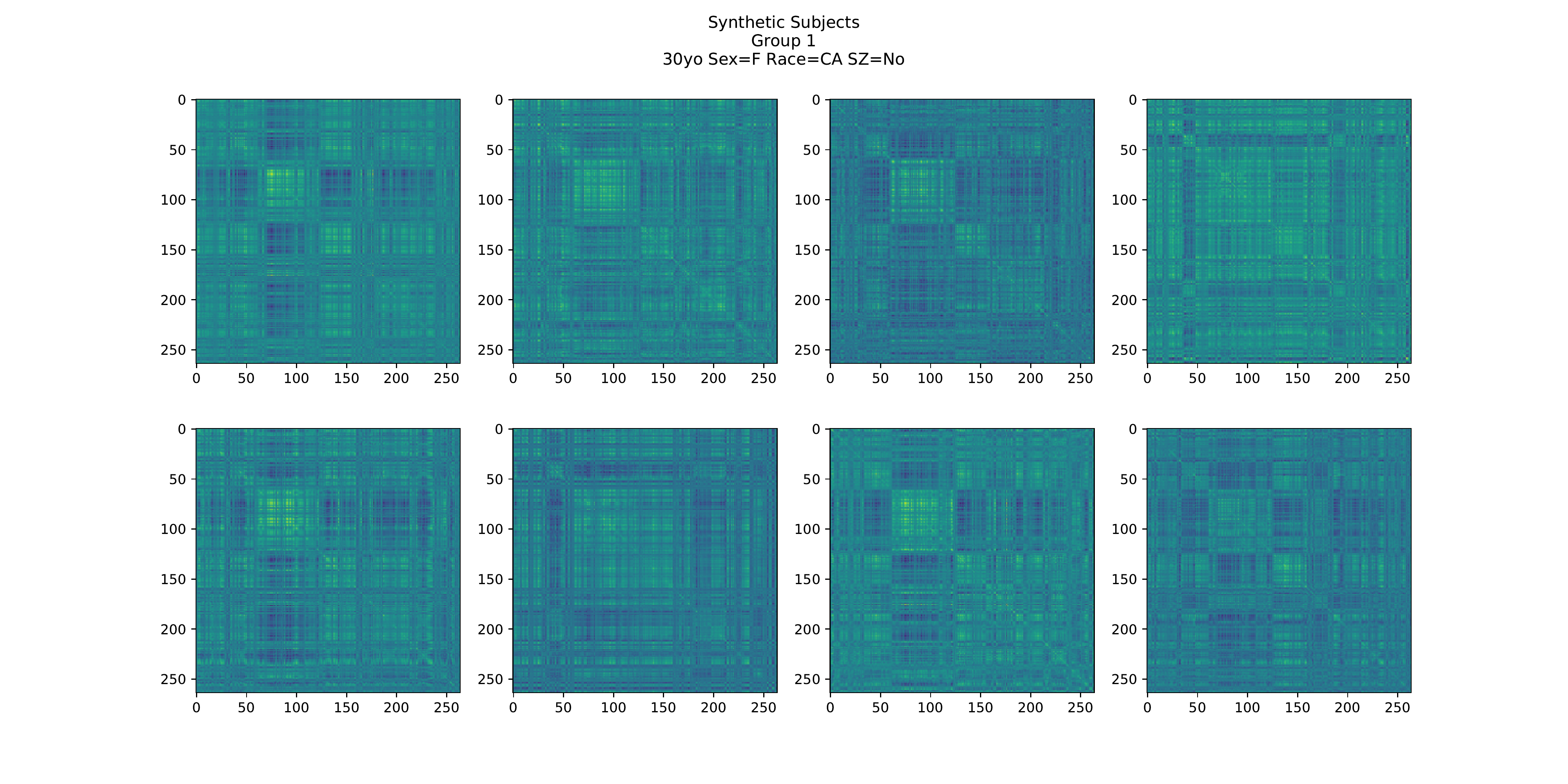}
    \hspace*{-1cm}
    \includegraphics[width=14cm]{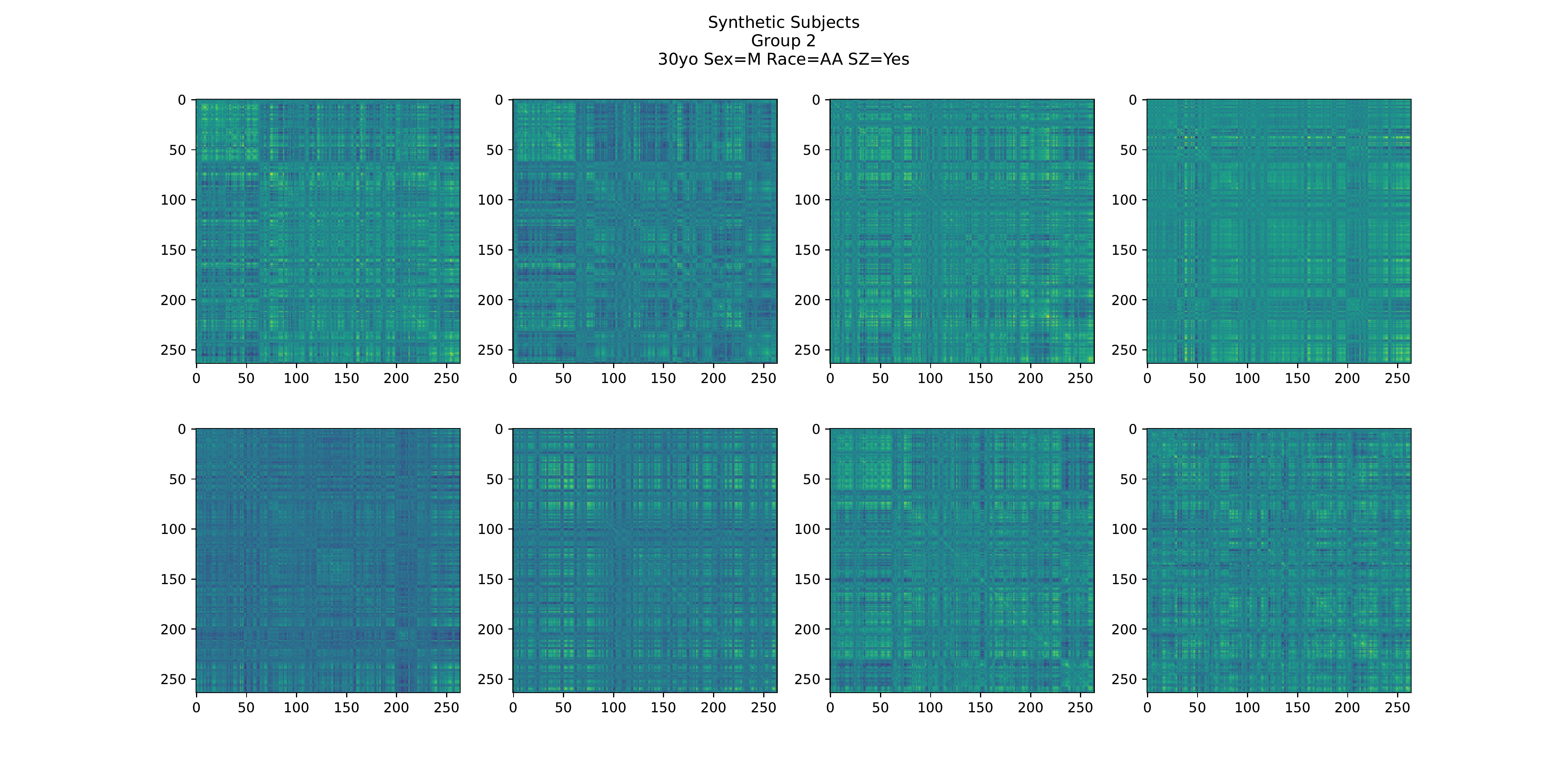}
    \caption{First 8 synthetic subjects from 30yo normal CA female group and 30yo schizophrenic AA male group.}
    \label{fig:groups}
\end{figure}

One hundred synthetic subjects were created by initializing an angle basis model with 5 bases to random values, then performing gradient descent on the loss between the FC reconstructed from the angle basis and the specified values. A stop gradient operation was applied to model weights, so only the angle basis values (and consequently FC) could change. The phenotype parameters were set to 30 years for age and $\pm10$ for the race, sex, and schizophrenia diagnosis logits. For reference, the absolute value of logits from the original predictive models ranged from 4 to 6 on their respective training sets. Gradient descent was performed for 100 epochs using the Adam optimizer with a learning rate of $10^{-2}$ and no L2 regularization. The first 8 synthetic FCs for both groups are shown in Figure~\ref{fig:groups}.

\begin{figure}
    \centering
    \hspace*{-1cm}
    \includegraphics[width=14cm]{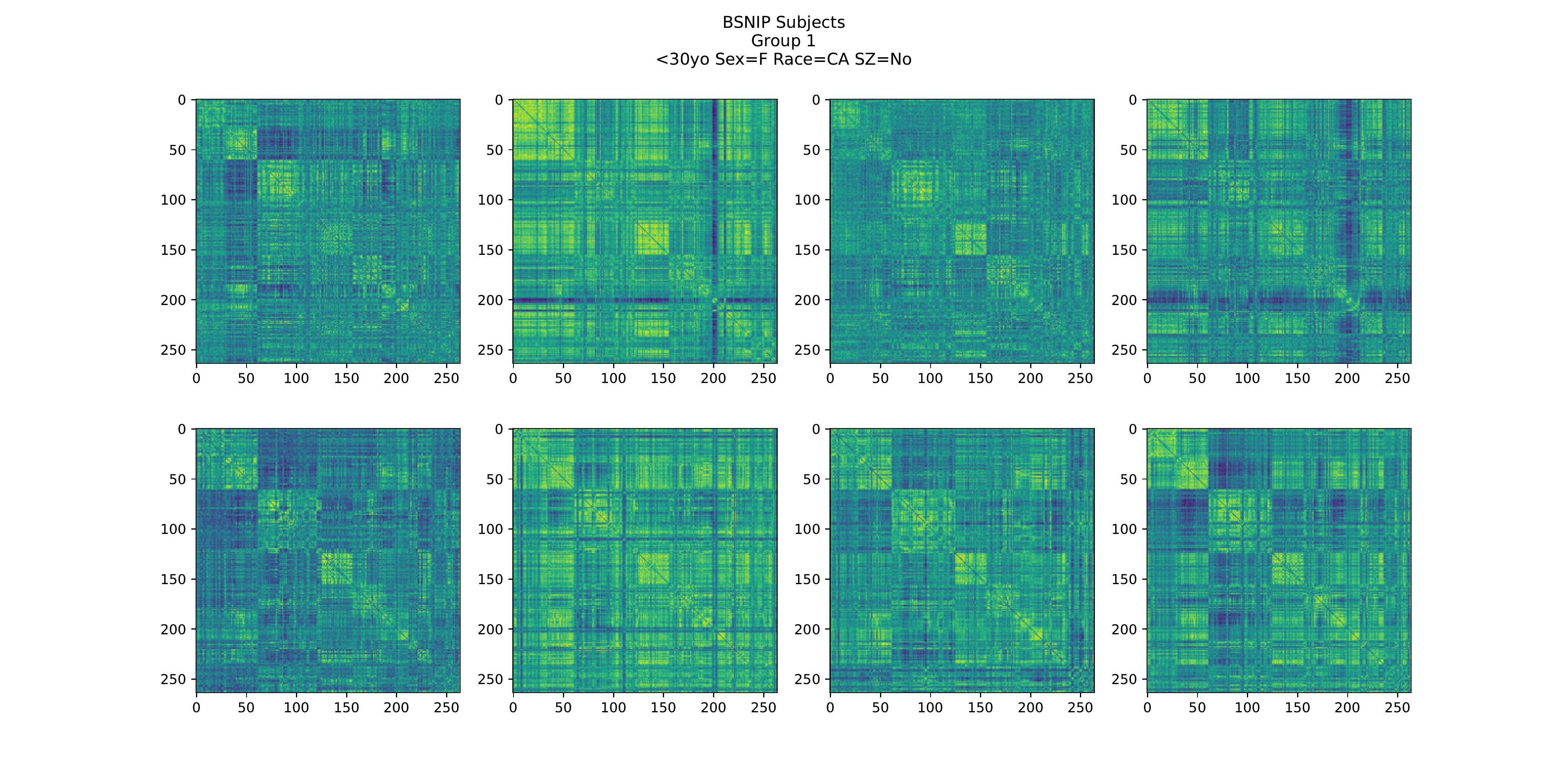}
    \hspace*{-1cm}
    \includegraphics[width=14cm]{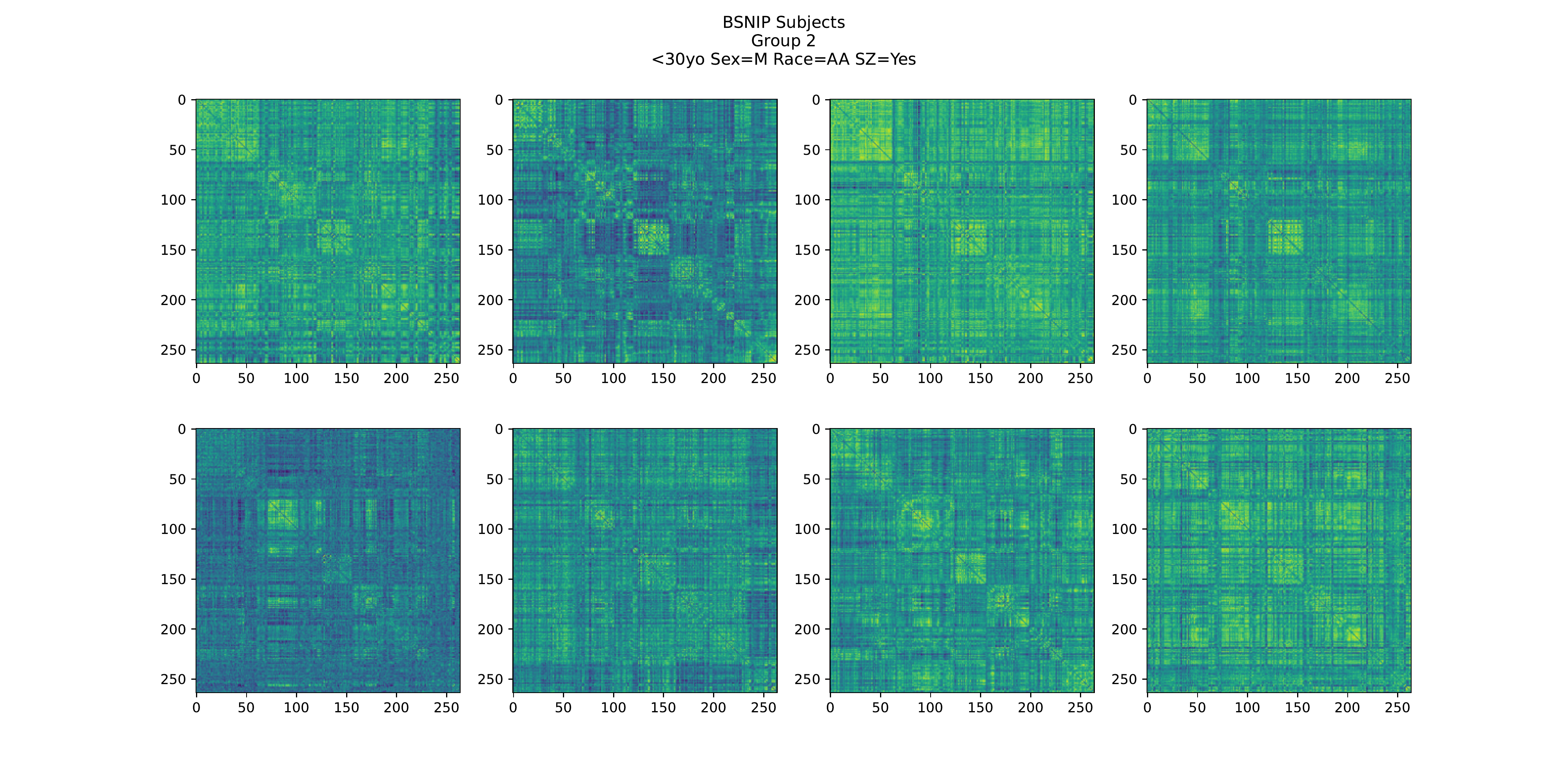}
    \caption{Example of the true variability of FC in the BSNIP dataset.}
    \label{fig:bsnip_groups}
\end{figure}

It should be noted that true subject FC exhibits far more variation than is captured in the synthetic data (see Figure~\ref{fig:bsnip_groups}). Nonetheless, we believe that the synthetic FC is useful since it accurately captures the coarse brain network-level differences between groups, and gives a simple visual description of qualitative differences in low-rank brain networks.  

\section{Prediction with Limited Number of Features}

\begin{figure}
    \centering
    \hspace*{-0.25cm}\includegraphics[width=6cm]{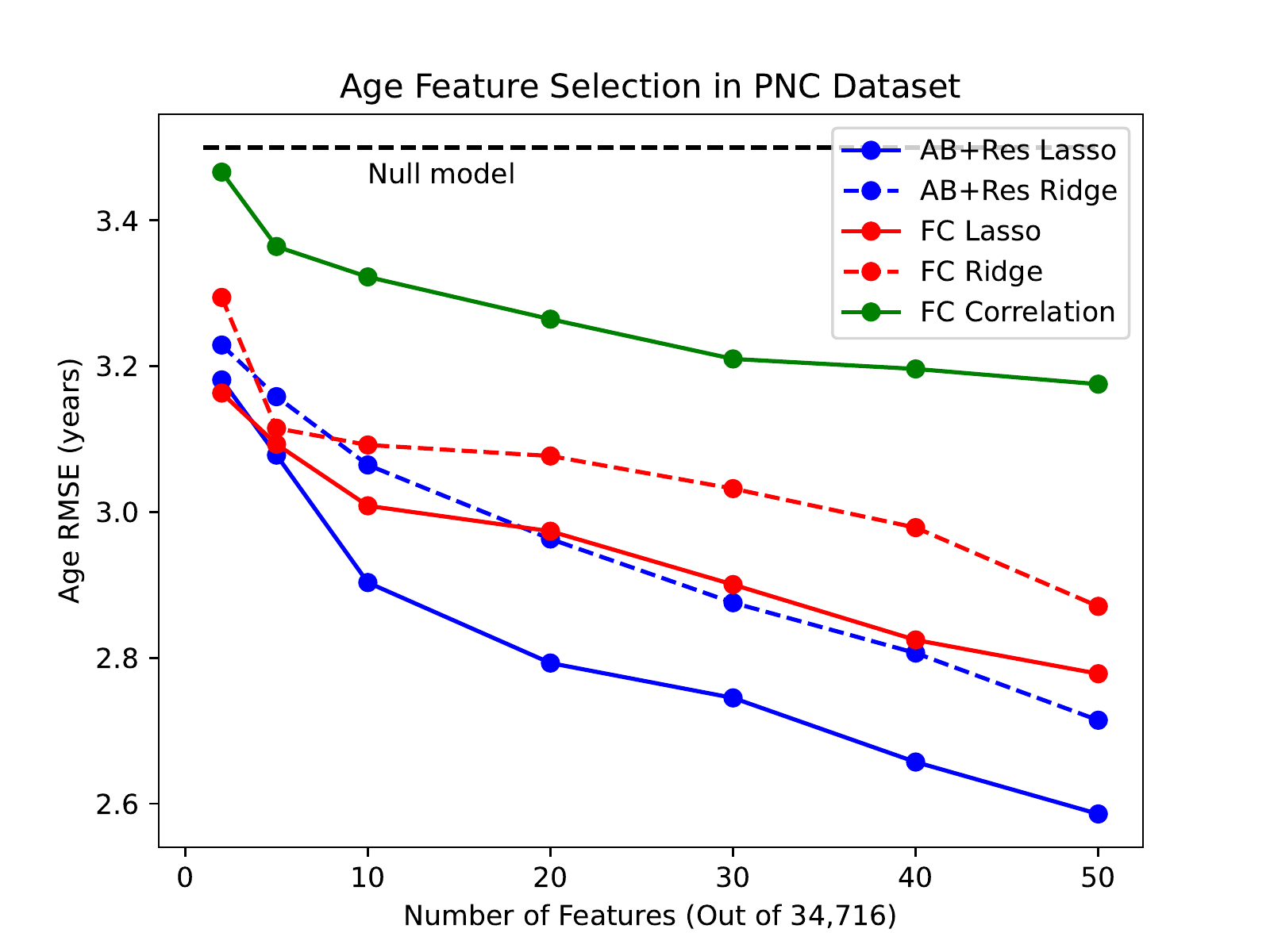}
    \caption{Feature selection for age prediction in the PNC dataset. Note features selected for angle basis reconstruction could be different than features selected for the angle basis residual.}
    \label{fig:features}
\end{figure}

We tested prediction on angle basis reconstruction and residual using features selected by Lasso, Ridge Regression, or correlation with the response variable. The results are shown in Figure~\ref{fig:features}. As expected, Lasso outperformed Ridge. Selecting features by correlation with the response variable was found to be a poor way to select features, likely due to the large multicollinearity present in FC.

\end{document}